\documentclass[preprint,showpacs,preprintnumbers]{revtex4}

\usepackage{amssymb}
\usepackage{amsfonts}
\usepackage{graphicx}
\usepackage{dcolumn}
\usepackage{bm}
\usepackage[dvipdfm]{hyperref}

\begin{document}

\title{Theory of field emission}
\author{Shi-Dong Liang}
\altaffiliation{Email: stslsd@mail.sysu.edu.cn}
\affiliation{School of Physics,\\
State Key Laboratory of Optoelectronic Material and Technology, and
Guangdong Province Key Laboratory of Display Material and Technology, Sun Yat-Sen University, Guangzhou, 510275, People's Republic of China}
\date{\today }

\begin{abstract}
A serious barrier impedes the comparison between the theoretical prediction and the
experimental observation in field emission because there is no way to measure the
emission area. We introduce three dimensionless variables $\mathcal{R}_{J}$, $\mathcal{S}$ and $\mathcal{D}$ to construct a formulation for connecting directly the theoretical variables and experimental data without measuring the emission area.
Based on this formulation we can analyze that the behaviors of $\mathcal{R}_{J}$
$\mathcal{S}$ and $\mathcal{D}$ with the voltages between the anode and the emitter to reveals the characteristics of current-voltage (I-V) curve and detect the physical properties of emitters. This formulation provides a way to understand the fundamental physics of I-V curve in field emission and to set up a map between the physical properties of emitters and
the experimental I-V curve.
\end{abstract}

\maketitle



\section{Introduction}
As a cold electron beam or source, field emission has been applied in
various vacuum nanoelectronics and its relevant equipments, such as such as
field emission microscope (FEM), electron
microscope (EM), transmission electron microscope (TEM), scanning electron
microscope (SEM), atom probe field ion
microscope (APFIM) and mass spacetrometer etc.. \cite{Xu} In recent years, with the
nanotechnology develops rapidly many new kinds of nanomaterials
are used in field emission, such as semiconducting nano wires and
carbon nanotubes, which opens two new chapters of field emission. One is to
develop some new field-emission-based devices, such
as the field-emission based lighting elements, the field-emission-based displays,
the field-emission-based sensor, the radio frequency devices, the field emission
information storage, the high-resolution imaging devices etc.\cite{Xu}
The other stimulates the theoretical interest how
to understand the field emission behavior of these new kinds of nano
materials because there have been experimental evidences, \cite{Bonard} which
indicate the current-voltage curves (I-V) violating the Fowler-Nordheim (FN) and
Schottky-Nordheim (SN) theories. \cite{Bonard,Gogolin,Liang1}
More interestingly, Some generalized formulations beyond the Fowler-Nordheim theory have been proposed for the Luttinger Liquids and the single-wall carbon nanotubes. \cite{Gogolin,Liang1} Understanding the behaviors of
field emission not only allows us to improve the performance of field
emission for these new kinds of nano materials, but also provide a way to
detect the physical properties of new nano-materials by field emission.\cite{Liu}

In fact, when we compare the theoretical prediction and experimental data we
face a fatal difficulty.\cite{Liang2} There is no a directly way to compare the
theoretical prediction and experimental data for all theoretical
formulations of field emission because all theoretical formulations of field
emission can be expressed only in terms of the current density versus the
local field at the emitter, $J(F)$, \cite{Gogolin,Liang1,Liu,Liang2,Gadzuk,Forbes1}
but the experimental measurements are only the total current versus the voltage between the anode and emitter.\cite{Xu,Bonard,Liu,Forbes1,Forbes2}
In general, the theoretical total current should be expressed as
\begin{equation}\label{I1}
I_{th}=\int_{\Omega }J(F)dA
\end{equation}
where $J(F)$\ is the current density and $F$ is the local field at emitter,
which depends on the position at the emitter, $F=F(A)$. However, there is no way to
measure the emission area and the local field at the emitter. The
experimental measurements are achieved only for the total current versus
the voltage between the anode and the emitter, namely $I_{exp }=I(V)$. The local field at emitter depends on the voltage between the anode and the emitter. One introduces the voltage conversion factor $\beta $\ to estimate the local field from the voltage between the anode and the emitter, $F=\beta V$.\cite{Forbes2,Forbes3} Hence, this fundamental difficulty has been blocking the connection between the theoretical prediction and the experimental measurements for almost nine decades. To solve this problem is
to propose a new theoretical formulation to connect the current density
directly to the total current and the voltage between the anode and the emitter,
in which we do not need to measure the emission area and local field.

The goal of this paper, we propose a new formulation to analyze directly the
current-voltage curve based on the experimental data without measuring emission area and local field at cathode. We introduce three dimensionless variables to
detect the physical properties of emitters based on the current-voltage curve directly.
In Section 2 we will present the generalized Schottky-Nordheim theory
as a unified theoretical model of field emission, which can describe various
materials of emitter in field emission. In Section 3
we will propose three dimensionless observables to construct a formulation for
connecting the theoretical variables to the experimental data directly. In Section 4,
we will analyze numerically the field emission for various materials of emitters.
Finally, we will give the conclusion.

\section{Generalized Schottky-Nordheim theory}
The first field emission model was proposed by
Fowler-Nordheim (FN) \cite{Fowler} based on the Summerfield's metallic model and quantum tunneling theory. Later, Schottky and Nordheim (SN) realized that the image potential effect
for the metallic emitters should enhance the field emission performance.\cite{Schottky}
Murphy Good developed this idea to formulate the image potential effect.\cite{Good}
They derive two correction factors to modify the FN theory based on the image potential effect.\cite{Liang2,Good,Forbes3,Forbes4} Further, the image potential effect was modified by introducing a parameter $\lambda$ to tune its strength for different semiconductors and rewrite the FN and SN formulas into a compact form,\cite{Liu,Dobretsov,Forbes5}
\begin{equation}\label{Jsn}
J(F,\lambda)=a\tau _{F}^{-2}(\lambda)\frac{F^{2}}{\phi }
\exp \left( -\nu _{F}(\lambda)\frac{b\phi^{3/2}}{F}\right)
\end{equation}
where J is the emission current density and F is the local field at the emitter.
$ a =\frac{e^{3}}{16\pi ^{2}\hbar }=1.5414 \mathrm{\mu A eV V^{-2}},
b =\frac{4}{3}\frac{(2m_{e})^{1/2}}{e\hbar }=6.83089 \mathrm{eV^{-3/2}V nm^{-1}} $
are the fundamental field emission parameters. $\phi $ is the local work function of emitter.
The correction factors $\tau_{F}$ and $\nu_{F}$ are modified,\cite{Liu,Dobretsov,Forbes5}
\begin{eqnarray}
\tau _{F}(\lambda) &\approx &1+\frac{\lambda}{9}f^{SN}-\frac{\lambda}{18}f^{SN}\ln (\lambda f^{SN})
\label{tau}\\
\nu _{F}(\lambda) &\approx &1-\lambda f^{SN}+\frac{\lambda}{6}f^{SN}\ln (\lambda f^{SN})
\label{nu}
\end{eqnarray}
where $f^{SN}=\frac{e^{3}}{4\pi \varepsilon _{0}\phi^{2}}F
=c_{\phi}\beta V$ with
$c_{\phi}=\frac{e^{3}}{4\pi \varepsilon _{0}\phi^{2}}$

It should be pointed out that
we introduce the $\lambda$-dependent the correction factors in Eqs.($\ref{tau}$)
and ($\ref{nu}$) to describe the image potential effect for different materials of emitters.
\cite{Liu}
(1) When $\lambda=1$
the generalized field emission formula in Eq. ($\ref{Jsn}$) is the pure SN formula, which
means the emitter being metallic.
(2) When $\lambda=0$
the image potential effect vanishes and the formula reduces to the FN formula.
(3) When $0\leq\lambda\leq0.5$ the formula describes the p-type semiconductor emitters.
(4) When $0.5\leq\lambda\leq 1$ the formula describes the n-type semiconductor emitters.
This generalized SN formula in Eq. ($\ref{Jsn}$) has been verified by some numerical studies and experimental evidences.\cite{Liu}
However, this formalism gives the emission current density versus the local field at emitter, which cannot compare directly to the experimental data, total emission
current and voltage between the anode and the emitter even though it can work as a qualitatively estimation.
This is a fundamental difficulty in the field emission theory
because one cannot measure directly the emission area.

\section{Map between J(F) to I(V)}
How do we compare directly the theoretical variables and the experimental observation?
The basic idea includes two key factors. One is to find out some variables connected directly to the experimental data without measuring the emission area.
The other is that these variables can reveal some physical properties of emitters
such that we can detect some material properties of emitter based on field emission
data. Thus, we propose a dimensionless formulation to connect the theoretical variables and
the experimental data without measuring the emission area.
We define the ratios based on the generalized SN theory to detect the physical information from field emission observations.
\begin{equation}\label{R1}
\mathcal{R}_{J}(F,\Delta F) := \frac{J(F+\Delta F)}{J(F)}
\end{equation}
where we ignore the parameter $\lambda$ for convenience and $\Delta F>0 $ is the small variation of the local field at anode.
The $\mathcal{R}_{J}(F,\Delta F)$ can be regarded as a dimensionless emission current.

Physically we can consider the following three features in field emission:
(1) As an approximation, the total emission current can be written as
$I(F)\approx J(F)A(F)$, where $J(F)$ is the local emission current density and
$A(F)$ is the emission area.
(2) For a small $\Delta F$, the change of the emission area is very small and disregarded,
$A(F)\approx A(F+\Delta F)$.
(3) The local field depends linearly on the voltage applied to anode,
$F=\beta V $, where $\beta$ is the voltage conversion factor, which can be determined by the experimental data. Hence, the variable $F$ in the expressions ($\ref{R1}$)
can be replaced by $V$,
namely $\mathcal{R}_{J}(F,\Delta F)=\mathcal{R}_{J}(V,\Delta V)$,
Thus, by using the formula in Eq.($\ref{Jsn}$) and notice that
$\frac{(F+\Delta F)^{2}}{F^{2}}=1+2\frac{\Delta V}{V}+\left(
\frac{\Delta V}{V}\right)^{2}\equiv \gamma_{v}$,
we can obtain
\begin{equation}\label{R2}
\mathcal{R}_{J}(V,\Delta V) =\gamma_{v}
\mathcal{S}_{th}^{-2}(V,\Delta V)\exp \left(-\mathcal{D}_{th}(V,\Delta V)\frac{b\phi ^{3/2}}
{\beta V } \right)
\end{equation}
where
\begin{eqnarray}\label{SD}
\mathcal{S}_{th}(V,\Delta V)  &=&
\frac{1+\frac{\lambda_{\Delta }}{9}-\frac{\lambda_{\Delta }}{18}\ln \lambda_{\Delta }}{1+\frac{\lambda_{V}}{9}-\frac{\lambda_{V}}{18}\ln \lambda_{V}} \label{S}\\ \nonumber
\mathcal{D}_{th}(V,\Delta V) &=&
\left(\frac{V}{V+\Delta V}\right)
\left(1-\lambda_{\Delta}-\frac{\lambda_{\Delta }}{6}\ln \lambda_{\Delta}\right)
\\
&-&\left(1-\lambda_{V}-\frac{\lambda_{V}}{6}\ln \lambda_{V}\right)\label{D}
\end{eqnarray}
with $\lambda_{\Delta}=c_{\phi}\lambda\beta(V+\Delta V)$ and $\lambda_{V}=c_{\phi}\lambda\beta V$. It should be noted that $\mathcal{S}_{th}(V,\Delta V)$ and $\mathcal{D}_{th}(V,\Delta V)$
depend on the physical parameters $\beta,\phi,\lambda$ and the voltage $V$ for given $\Delta V$, which can be called as the feature functions of field emission.
The subscript $th$ in $\mathcal{S}_{th}(V,\Delta V)$ and $\mathcal{D}_{th}(V,\Delta V)$  represents their theoretical feature of these expression. For small $\Delta V$, we have
\begin{equation}\label{JI}
\mathcal{R}_{J}(V,\Delta V)=\frac{J(V+\Delta V)}{J(V)}=\frac{I(V+\Delta V)}{I(V)}\equiv R_{I}(V,\Delta V)
\end{equation}
This is the map between the theoretical model ($R_{J}$) and the experimental data
($R_{I}$) directly without measuring the emission area.
Thus, we can rewrite the expressions of $\mathcal{S}_{th}(V,\Delta V)$ and
$\mathcal{D}_{th}(V,\Delta V)$ to $\mathcal{S}_{exp}(V,\Delta V)$ and
$\mathcal{D}_{exp}(V,\Delta V)$ in terms of $R_{I}$,
\begin{eqnarray}\label{SD2}
\mathcal{S}_{exp}(V,\Delta V)&=&\left[\frac{\gamma_{v}}{R_{I}(V,\Delta V)}\right]^{1/2}
\exp\left(-\mathcal{D}_{th}(V,\Delta V)\frac{b\phi ^{3/2}}
{2\beta V } \right) \\
\mathcal{D}_{exp}(V,\Delta V) &=& -\frac{\beta V}
{b\phi ^{3/2} } \ln \left(\frac{R_{I}(V,\Delta V) \mathcal{S}_{th}^{2}(V,\Delta V)}{\gamma_{v}}\right)
\end{eqnarray}
It should be remarked that $\mathcal{S}_{exp}(V,\Delta V)$ and
$\mathcal{D}_{exp}(V,\Delta V)$ depend on the experimental data because they involve $R_{I}(V,\Delta V)$.
Consequently, these three theoretical variables $\mathcal{R}_{J}$, $\mathcal{S}_{th}(V,\Delta V)$ and $\mathcal{D}_{th}(V,\Delta V)$ we introduce provide a way
to compare directly with the experimental data $\mathcal{R}_{I}$, $\mathcal{S}_{exp}(V,\Delta V)$ and $\mathcal{D}_{exp}(V,\Delta V)$ based on the generalized Schottky-Nordheim model. By comparison between the theoretical
variables and the experimental data from the I-V curve we can detect
the physical parameters $\phi$ and $\lambda$ to reveal the physical properties and
the basic physics of field emission.

In practice, $\Delta V$ depends on the steps of the experimental measurements.
In principle, the voltage conversion factor $\beta $ can be estimated
by the experimental data from the I-V curve. We may consider
that $\lambda=0$ as an estimation of $\beta$, namely $\mathcal{S}(V,\Delta V)=1$ and
$\mathcal{D}(V,\Delta V)=-\frac{\Delta V}{V+\Delta V}$, we define
\begin{equation}\label{y}
y:=\ln\left[\frac{R_{I}(V,\Delta V)}{\gamma_{v}}\right]
\end{equation}
and
\begin{equation}\label{x}
x:=\frac{\Delta V}{V(V+\Delta V)}
\end{equation}
where $y$ and $x$ depend only on the voltage and
the I-V curve from the experimental data.
The expression of the ratio in Eq. ($\ref{R2}$) for $\lambda=0$ can be rewritten as
\begin{equation}\label{y2}
y =-\frac{b\phi ^{3/2}x}{\beta}
\end{equation}
We can plot the slope of $y\sim x$, namely $S_{\beta}:=\frac{dy}{dx}$ based on the I-V curve from the experimental data for given $\Delta V$. Thus,
the voltage conversion factor $\beta$ can be obtained approximately by
\begin{equation}\label{beta}
\beta=-\frac{b\phi ^{3/2}}{S_{\beta}}
\end{equation}
which is equivalent to the conventional FN-plot method.\cite{Forbes2}
However, here we use $R_{I}$ instead of $J(F)$ without measuring emission area.

\begin{figure}
\resizebox{1\hsize}{!}{\includegraphics*{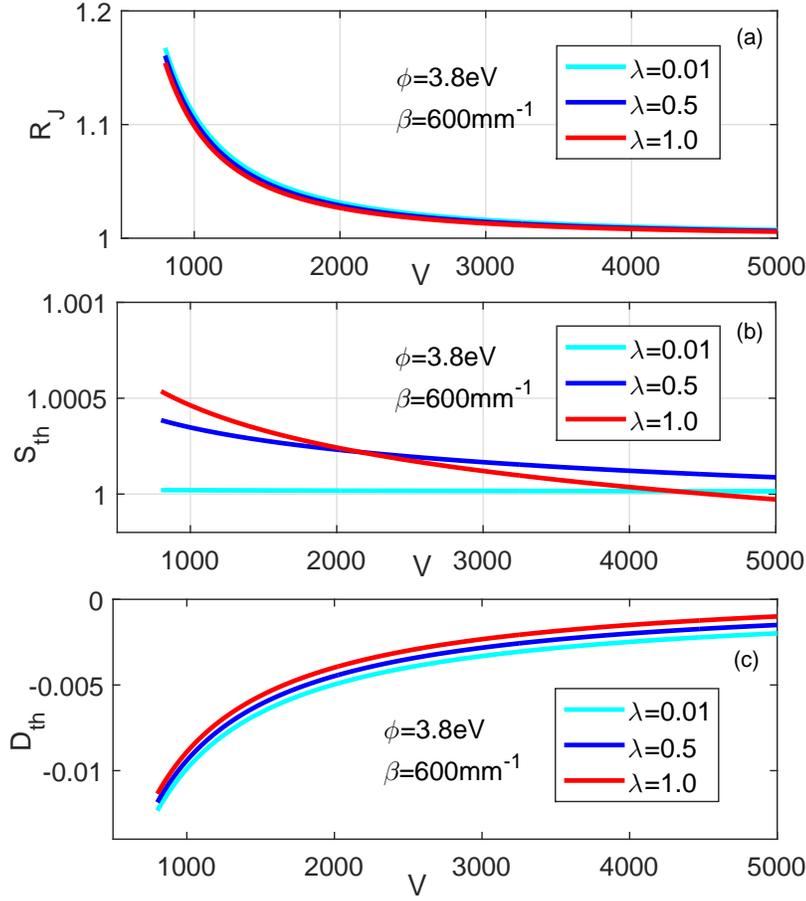}}
\caption{\label{Fig1}
(Color online): The signals of field emission for different parameters $\lambda$.}
\end{figure}

The theoretical formulation from ($\ref{R1}$) to ($\ref{beta}$) provides a
map between the theoretical model and experimental data
without measuring the emission area.
The local work function $\phi$ and $\lambda$ depend
on the metallic or semiconducting properties.\cite{Liu}
The voltage conversion factor is related to
the distance between the emitter and anode as well as the shape of emitter,
but it can be estimated by the experimental plot from Eq. ($\ref{beta}$).
Theoretically we can analyze the basic behaviors of these variables
to reveal the material properties of emitters based on this formulation.

\begin{figure}
\resizebox{1\hsize}{!}{\includegraphics*{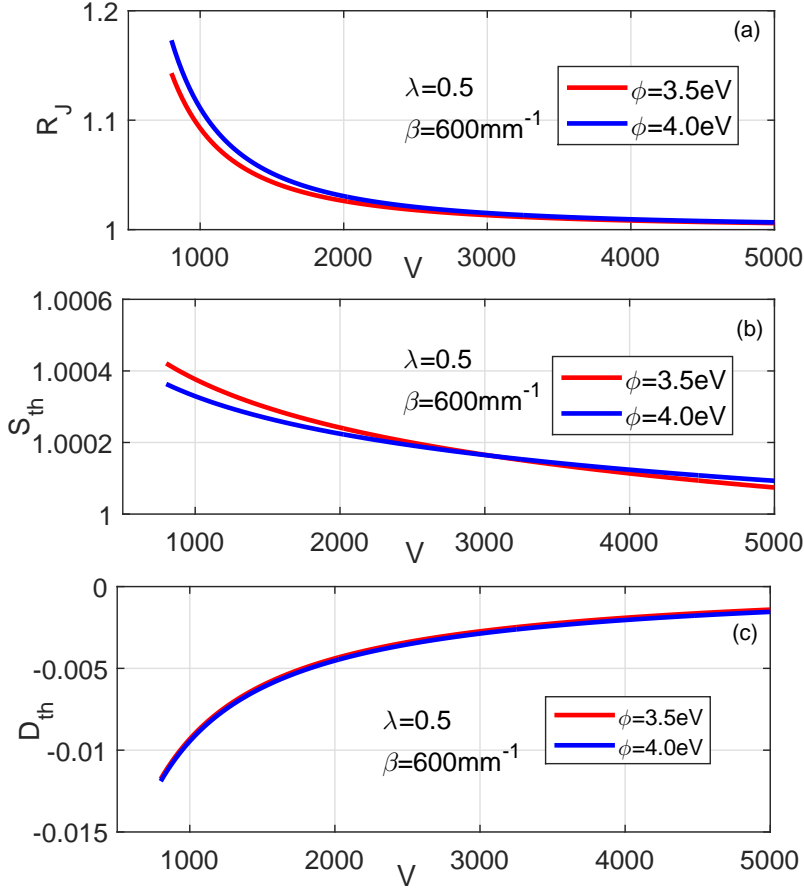}}
\caption{\label{Fig2}
(Color online): The signals of field emission for different local work functions.}
\end{figure}

For given small $\Delta V$, the formulations of $\mathcal{R}_{J}$, $\mathcal{S}_{th}$ and $\mathcal{D}_{th}$ depend only on $V$. In principle,
as the theoretical analysis to field emission,
we can also investigate $\mathcal{R}_{J}$, $\mathcal{S}_{th}$ and $\mathcal{D}_{th}$
versus $V$ for different small $\Delta V$. Nevertheless, the variance of $\Delta V$
in a small range does not change the basic behaviors of $\mathcal{R}_{J}$, $\mathcal{S}_{th}$ and $\mathcal{D}_{th}$ with $V$. In the practical experiments, we can make the
averages of $\mathcal{R}_{J}$, $\mathcal{S}_{th}$ and $\mathcal{D}_{th}$ for different $\Delta V$ to reduce the errors from variant $\Delta V$.

\begin{figure}
\resizebox{1\hsize}{!}{\includegraphics*{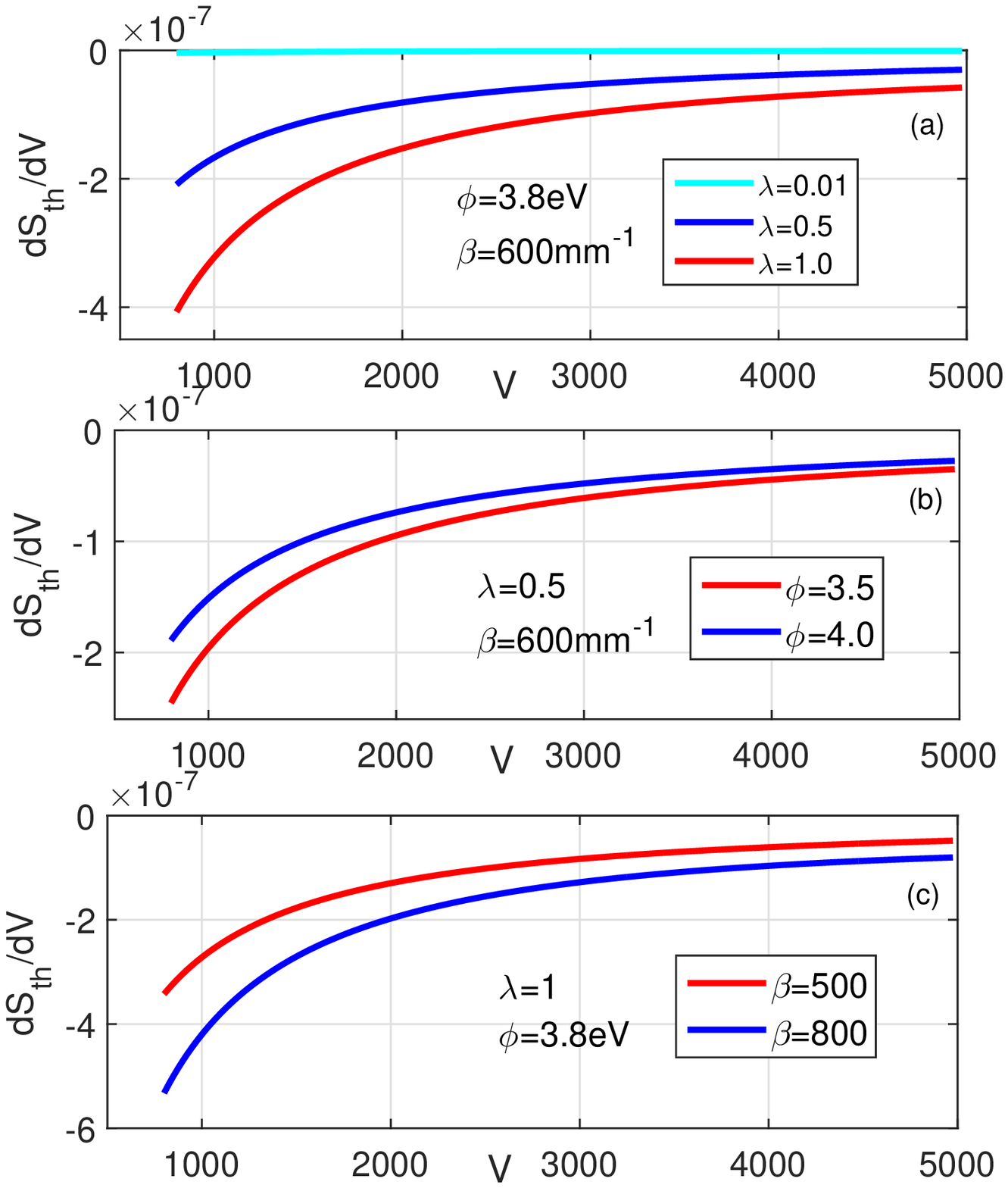}}
\caption{\label{Fig3}
(Color online):The signals of field emission for different voltage conversion factors.}
\end{figure}

It should be remarked that the experimental setup of field emission usually involves some electric circuits, which could modify a little bit of the emission current and the voltage
between the anode and emitter. R. Forbes analyze and discuss this effect in detail.\cite{Forbes5} In principle the circuit effect can be taken into account
by the practical experimental setup. This is another issue, which does not change the basic idea of this formulation.

\section{Physical properties from I-V curve}
The numerical study of this formulation allows us to capture the basic physics of field emission. In general, the typical local fields of field emission are about
$10^{9}\sim 10^{10}\mathrm{V/M}$. Suppose that the voltages between the anode and emitter
are about $500\sim 5000 \mathrm{V}$, which are easy to be implemented in experiments,
thus, the voltage conversion factors are about $600\sim 800 \mathrm{mm^{-1}}$
such that the local field are within the typical field emission range.
We set $\Delta V=10\mathrm{V}$ for all figures, which does not change qualitatively the results.

Fig.$\ref{Fig1}$ shows that the behaviors of the ratio of the emission current $\mathcal{R}_{J}$
and the feature functions $\mathcal{S}_{th}$ and $\mathcal{D}_{th}$ with different $\lambda$ for given $\phi$ and $\beta$.
It can be seen that $R_{J}$ decreases with $V$ which depends a little on $\lambda$
and $\mathcal{D}_{th}$ increases with $V$ and is not sensitive to $\lambda$.
Interestingly the slopes of $\mathcal{S}_{th}$ depend on $\lambda$, which provides
a way to detect the physical properties of emitters.

We investigate the behaviors of $\mathcal{R}_{J}$, $\mathcal{S}_{th}$ and $\mathcal{D}_{th}$
with the local work function $\phi$ in Fig. $\ref{Fig2}$ for given $\lambda$ and $\beta$.
The behaviors of $\mathcal{R}_{J}$ $\mathcal{S}_{th}$ and $\mathcal{D}_{th}$ are
similar to that in Fig. $\ref{Fig1}$.

In Fig.$\ref{Fig3}$ we plot the derivatives of $\mathcal{S}_{th}$ with $V$ for different $\lambda$, $\phi$ and $\beta$. It can be seen that $\frac{d\mathcal{S}_{th}}{dV}$ depend
on $\lambda$, $\phi$ and $\beta$, which provides a way to understand the physical
properties of emitters because the physical properties or the geometric shapes of emitters, such as metallic and n- or p-type semiconductors as well as nanowires, depend on $\lambda$,
$\phi$ and $\beta$.\cite{Liu}

Fig. $\ref{Fig3}a$ indicates that when $\lambda\rightarrow 0$, $\frac{d \mathcal{S}_{th}}{dV}\rightarrow 0$ and
$\left|\frac{d \mathcal{S}_{th}}{dV}\right|$ increases with
$\lambda$, which reflects the physical properties of emitters. Similarly in Figs. $\ref{Fig3}b$ and $c$, the $\phi-$ and $\beta-$dependent behaviors of $\frac{d \mathcal{S}_{th}}{dV}$ reveal the basic physics of I-V curve.

It should be remarked that $\mathcal{R}_{J}$, $\mathcal{S}_{th}$ and $\mathcal{D}_{th}$
we introduce are dimensionless, which can identify the intrinsic physical properties
of I-V curve. Especially $\frac{d \mathcal{S}_{th}}{dV}$
provides a way to analyze or understand the basic physics of I-V curve in field emission
even though their values are small.

\section{Conclusions}
In summary, we introduce three dimensionless variables $\mathcal{R}_{J}$, $\mathcal{S}$ and $\mathcal{D}$ to construct a formulation for connecting directly the theoretical model
and experimental data without measuring the emission area.
Based on this formulation, we find that the slopes of $\mathcal{S}$
depend on the physical parameters, $\lambda$, $\phi$ and $\beta$, which
reveals the physical characteristics of I-V curve and their relationships
to the physical properties of emitters, such as metallic and n- or p-type semiconductors
as well as nanowires. This formulation provides a way to understand
the fundamental physics of I-V curve in field emission and
can be expected to apply for setting up a map between the physical properties of emitters and the experimental I-V curve.
Understanding the basic physics of I-V curve helps us to improve the field emission performance by designing the material and structure of
emitters from tuning the parameters of $\phi$, $\beta$ and $\lambda$.
On the other hand, we can also apply this formulation to test or detect
the physical properties of new kinds of materials by
fitting the I-V curve to obtain the material parameters, $\lambda$, $\phi$ and $\beta$ .\cite{Liu}

\begin{acknowledgements}
The authors gratefully thank the the Natural Science Foundation of Guangdong Province (No. 2016A030313313).
\end{acknowledgements}





%

\end{document}